\documentclass[journal]{IEEEtran}
\usepackage{amsmath,amssymb,amsfonts}
\usepackage{algorithmic}
\usepackage{algorithm}
 \usepackage{array}
\usepackage{stackengine}
 \usepackage{textcomp}
 \usepackage{stfloats}
 \usepackage{url}
 \usepackage{verbatim}
 \usepackage{graphicx}
  \usepackage{multirow}
 \usepackage{subcaption}
 \usepackage{xcolor}
 \usepackage{cite}

\begin{document}

\title{Analysis of LGM Model for sEMG Signals related to Weight Training}
\author{Durgesh Kusuru, \IEEEmembership{Student Member, IEEE}, Anish C. Turlapaty \IEEEmembership{Member, IEEE} and \\ Mainak Thakur  \IEEEmembership{Member, IEEE}
\thanks{The authors are with the Bio-signal Analysis Group, Indian Institute of Information 
Technology Sri City, Chittoor, Andhra Pradesh, 517646, India. 
(e-mails: durgesh.k@iiits.in, anish.turlapaty@iiits.in and mainak.thakur@iiits.in}}


\markboth{Submitted to IEEE TIM}%
{Kusuur \MakeLowercase{\textit{et al.}}: MODEL-ANALYSIS-EMAHA-DB2}


\maketitle

\begin{abstract}
\textcolor{black}{
Statistical models of Surface electromyography (sEMG)  signals have several applications such as better understanding of sEMG signal generation, improved pattern recognition based control of wearable exoskeletons and prostheses, improving training strategies in sports activities, and EMG simulation studies.  
Most of the existing studies analysed the statistical model of sEMG signals acquired under isometric contractions. However, there is no study that addresses the statistical model under isotonic contractions. In this work, a new dataset, electromyography analysis of human activities - database $2$ (EMAHA-DB2) is developed. It consists of two  experiments based on both isometric and isotonic activities during weight training. {Previously, a novel Laplacian-Gaussian Mixture (LGM) model was demonstrated for a few benchmark datasets consisting of basic movements and gestures. }In this work, the model suitability analysis is extended to the EMAHA-DB2 dataset. Further, the LGM model is compared with three existing statistical models including the recent scale-mixture model. According to qualitative and quantitative analyses, the LGM model has a better fit to the empirical pdf of the recorded sEMG signals compared with the scale mixture model and the other standard models. The variance and  mixing weight of the Laplacian component of the signal are analyzed with respect to the type of muscle, type of muscle contraction, dumb-bell weight and training experience of the subjects. {The sEMG variance (the Laplacian component) increases with respect to the weights, is greater for isotonic activity especially for the biceps. For isotonic activity, the signal variance increases with training experience. Importantly, the ratio of the variances from the two muscle sites is observed to be nearly independent of the lifted weight and consistently increases with the training experience.}
}
\end{abstract}
\begin{IEEEkeywords}
Surface electromyography (sEMG), Statistical model, Laplacian-Gaussian Mixture (LGM) model, EMAHA-DB2, Scale-mixture  model, Isotonic, Isometric and Muscle contraction force.
\end{IEEEkeywords}
\section{Introduction}
\subsection{Background}
\textcolor{black}{
\IEEEPARstart{S}{tatistical} modeling of the strength of surface  Electromyography (sEMG) signals is an important problem with  applications in EMG based movement classification \cite{fleischer2006application, fukuda2003human, furui2021emg}, understanding EMG signal generation, EMG signal simulation, clinical interpretation, biomechanics, and visualization for movement sciences. {Refer \cite{kusuru2023laplacian} and the references therein for an extensive motivation for the EMG signal modeling.} } 

\textcolor{black}{
There are several studies that model's EMG signals from various arm muscles under different conditions. According to the following studies, \cite{nazarpour2005negentropy,nazarpour2013note,naik2011evaluation,naik2011kurtosis,thongpanja2013probability,furui2019scale,milner1975relation}, sEMG signals recorded from the muscle sites on the forearm, biceps, and triceps under constant force, constant angle, and non-fatigue contractions follow the Laplacian and the Gaussian models at low and  high contraction levels respectively. However based on a few other studies \cite{hussain2009electromyography,kaplanis2000bispectral} , the EMG signals follow the Laplacian model at higher contraction levels.}
\textcolor{black}{
In our previous work \cite{kusuru2023laplacian} and \cite{kusuru2021laplacian}, we have proposed a Laplacian Gaussian mixture (LGM) model for sEMG signals obtained from upper limbs. The LGM merges both Gaussian and non-Gaussian components into a single model. The model was validated with standard Gaussian and Laplacian models. It was shown that LGM model is superior to standalone Gaussian and Laplacian models. In this paper, the suitability of the LGM model is analyzed for sEMG signals acquired from arm muscles during exercise activities under various loading conditions.}
 
\subsection{Problem statement}

\textcolor{black}{ Based on existing literature, sEMG signal strength depends on type of muscle and muscle contraction force. It is also influenced by several other factors such as electrode shift, limb positions, dynamic arm movements, uncertain force disturbances, muscle fatigue, inter-subject variability, ethnicity \cite{doud1995muscle,gallagher1997appendicular,young2011effects}, nutrition and medium of measurements \cite{gallagher1997appendicular,young2011effects,da2010analysis,kyranou2018causes,yang2015classification}. The focus of this paper is to determine a suitable model for signal strength of sEMG signals from arm muscles during different exercise activities. Previous studies have examined the statistical nature of EMG signals under isometric activities \cite{furui2019scale,thongpanja2013probability,clancy1999probability}. However, there is no study on the statistical model of sEMG signals under isotonic contractions. Hence in this paper the statistical model and its properties are investigated for both isomteric and isotonic contractions.} 

\textcolor{black}{ Weight training is effective for improving muscle strength, overall health, and regaining limb functionality for people undergoing rehabilitation post stroke-related episodes. The EMG signals acquired during weight training can be used for muscle recruitment analysis. For example, during a specific movement, it can determine the set of recruited muscles and their order of recruitment \cite{singh2019time}. In this study, for a set of subjects from the Indian population, sEMG signals are acquired during weight training activities involving both  isometric and isotonic contractions. There are several muscles involved in these activities, including those within the wrist, the forearm, biceps, and triceps. In this work, the focus is on the biceps and forearm muscles. Specifically, the Biceps brachii (BB)  and  Flexor carpi ulnaris (FCU) muscles are considered. The objective is to understand the variations in the statistical properties as a function of lifting weights and different measurement conditions.}

\subsection{Contributions}
\textcolor{black}{
The major contributions of the study are as follows }
\begin{itemize}

    \item \textcolor{black}{ A new surface EMG dataset is collected from subjects of the Indian population during weight training activities including isometric and isotonic contractions under various load conditions. }
    \item 
    \textcolor{black}{The suitability of the LGM model is analyzed for the acquired sEMG signals.}
    \item \textcolor{black}{The LGM model is validated through comparisons with the state of art \cite{furui2019scale} and the classical models.}
    \item  \textcolor{black}{The impact of the load on the sEMG signal model is analyzed. Specifically, the impact on the nature of the model, its composition and the specific model parameters is analyzed.} 
    \item \textcolor{black}{Finally, the variations in the model statistics are analyzed with respect to other measurement parameters such as the type of muscle, weight training experience of a subject, and the type of activity.
    }
\end{itemize}

The rest of the work is structured as follows. The Laplacian-Gaussian Mixture (LGM) model structure is described in Section-II. Section-III explains the experimental setup for the model validation, followed by section-IV of these experiments, and Section-V gives discussion. Finally, Section VI concludes this study.

\section{Materials and Methods} 
\subsection{LGM MODEL OF SURFACE EMG SIGNALS}
\begin{figure}[!t]
        \centering
      \includegraphics[width=1.05\columnwidth]{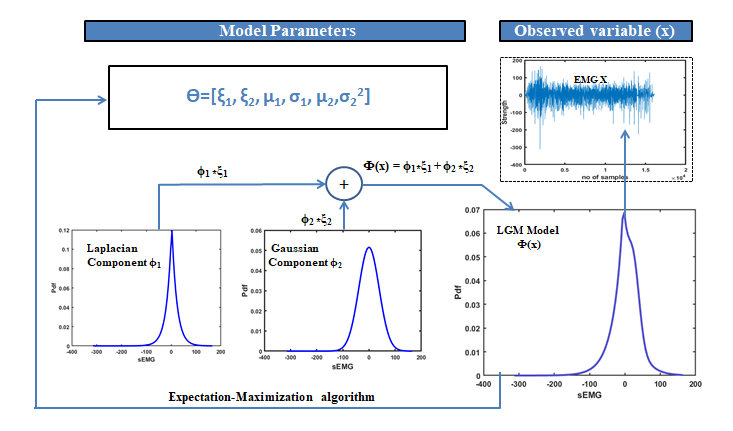}
      \caption{Graphical representation of LGM model: In this model, the sEMG signal $Z(n)$ is LGM random signal. The parameters of the model are estimated using EM-Algorithm. }
     \label{ms}
\end{figure}

\textcolor{black}{
Fig.\ref{ms} shows a graphical representation of the LGM model for sEMG signals. Let $Z(n)$ be random variable that specifies the values $z$ of EMG signal.
The LGM model can be written as follows
\begin{equation}
    \Phi(z;\Theta )= \lambda_{1}\phi_1(z;\theta_{1})+\lambda_{2}\phi_2(z;\theta_{2})
\end{equation}
 The strength of each component is determined by their corresponding weight coefficients ($\lambda_{1}$, $\lambda_{2}$).
$\phi_1(x;\theta_{1})$ and $\phi_2(x;\theta_{2})$ are the Laplacian and the Gaussian densities respectively. $\theta_{1}= (\mu_1,\sigma_1)$ and $\theta_{2}= (\mu_2,\sigma^2_{2})$ are parameters of the Laplacian and Gaussian densities. In this model, $\lambda_{1}$ and  $\lambda_{2}$ are interpreted as hidden variables. The vector $\Theta=[\lambda_1,\lambda_2,\theta_1,\theta_2]$ are unknown parameters and need to be estimated. In our previous work\cite{kusuru2021laplacian,kusuru2023laplacian}, the estimates for the vector $\theta$ were derived using the expectation-maximization algorithm and is implemented here.}

\subsection{Data Collection}
\begin{figure*}[t]
\captionsetup[subfigure]{justification=centering}
    \centering
      \begin{subfigure}{0.31\textwidth}
        \includegraphics[width=\textwidth]{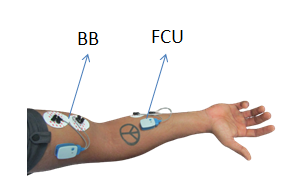}
          \caption{}
          \label{electrode placement}
      \end{subfigure}
      \begin{subfigure}{0.20\textwidth}
        \includegraphics[width=\textwidth]{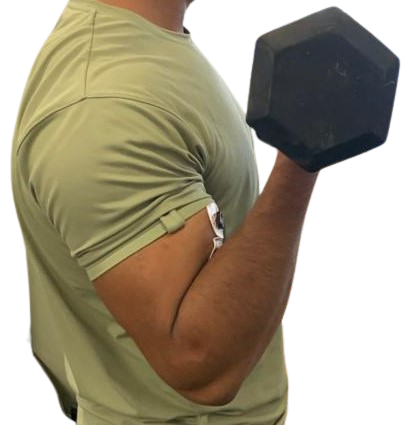}
          \caption{}
          \label{TONIC}
      \end{subfigure}
      \begin{subfigure}{0.26\textwidth}
        \includegraphics[width=\textwidth]{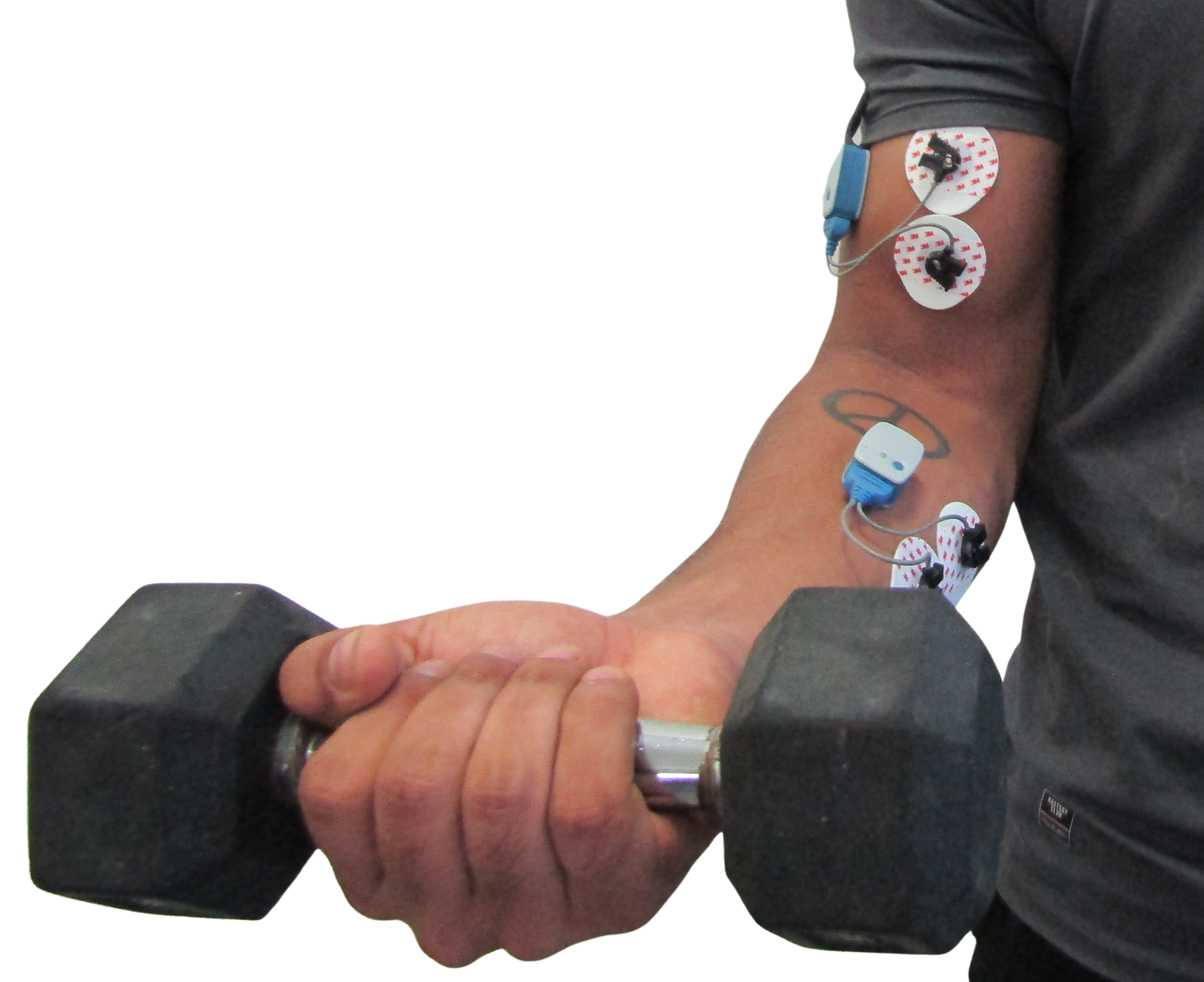}
          \caption{}
          \label{METRIC}
      \end{subfigure}
    
      \caption {(a) Placement of electrodes  on the BB and FCU muscles during dumbbell related activities. (b) Isotonic activity: Performing bicep curl exercise (c) Isometric activity: Holding the dumbbell weight at 90 $^{\circ}$  of flexion.  }
      \label{setup}
\end{figure*}

\textcolor{black}{
In this work, we develop a sEMG dataset termed electromyographic analysis of human arm activities - database 2 (EMAHA-DB2). Nine healthy participants aged between $18-21$ years were  selected on a voluntary basis. The subjects were selected based on three levels of weight training experience: a) Novice - with no prior weight training experience, b) Intermediate - with a few weeks of training and c) Trained - with at least one year of training. Participants were free from all muscle disorders for past one month prior to the day of the data collection. Prior to participating in the experiment, the purpose of the study was explained and an informed consent was obtained from the subjects. The data collection procedure was approved by the institutional ethics committee of the Indian Institute of Information Technology Sri City (No. IIITS/EC/2022/01). 
Before data acquisition process, the surface of the skin at the muscle site under consideration is cleaned with an alcohol based wipe to reduce the impedance. In EMAHA-DB2, sEMG signals are acquired using the Noraxon's Ultium sensors. As shown in fig. \ref{setup}(a), Ultium sensors are placed at two muscle sites 1) Biceps Brachii (BB) representing the upper arm activity and 2) Flexor carpi ulnaris (FCU) representing the forearm muscle activity. Signal acquisition characteristics of the sensor are: 16- bit A/D; Sampling rate: 2000 samples/sec; cutoff frequency: 20-450 Hz. The weights used during the activity include $0kg, 1kg ,2.5kg, 5kg, 6kg, 9kg$ and $10kg$.  During an activity, the subject is in a standing position and the weight is placed on a table at a convenient height. Each activity has three phases  1) rest (10s), 2) action (5s) and 3) release (3s) with a total duration of 18s. Each activity is repeated nine times. In order to avoid muscle fatigue, subjects rest for two minutes between different activities. Further details of experiments are given below. A summary of the dataset is presented in table \ref{tab:EMAHA-DB2}.}

 \subsubsection{Experiment-1}
 \textcolor{black}{
In the first experiment as shown in fig. \ref{setup}(b), the subjects were asked to perform bicep curls with the right arm using the seven weights mentioned above.      
Recall that the biceps curl corresponds to isotonic muscle contractions \cite{mayhew1995muscular}. } 
\subsubsection{Experiment-2}
\textcolor{black}{
In this experiment as shown in fig \ref{setup}(c), the subjects were asked to hold a dumbbell with their right hand at 90$^{\circ}$ with respect to the upper arm i.e., the dumbbell is held in the transverse plane with its axis parallel to the frontal axis. The same set of weight variations from experiment $1$ are used.     
{Recall, while holding a weight, the flexion of arm muscles corresponds to isometric contractions \cite{baley1966effects}.} } 

\begin{figure}[ht]
        \centering
      \includegraphics[width=.5\columnwidth]{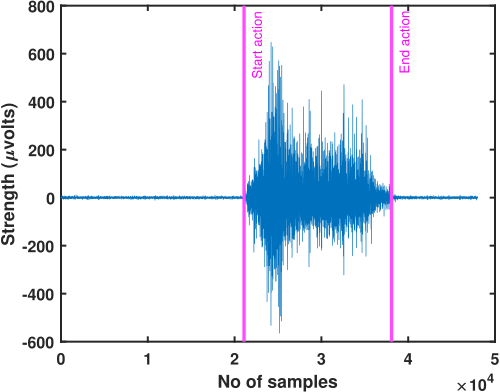}
      \caption{A sample raw-sEMG signal with the activity component segmentated}
     \label{R_S_DATA}
\end{figure}

\textcolor{black}{
The raw sEMG data is a mutually exclusive combination of rest and activity data. Prior to analysis of the data, the action component is segmented. \textcolor{black}{Segmentation is manually done by examining a raw signal directly and looking for the starting and ending points of the action.} Fig.\ref{R_S_DATA} shows a sample of raw sEMG signal (blue) and a segmented action component. The start and the end times of the action components are marked with vertical lines (purple). The segmented action components are subjected to the statistical analysis. The anthropometric details of the participants are shown in the table  \ref{details_participants}.}

\begin{table}[htb]
\centering
\caption{Characteristics of EMAHA-DB2 dataset}
\resizebox{\columnwidth}{!}{
\begin{tabular}{cccccccc}
\hline \hline
Name of activity       & \multicolumn{7}{c}{Dumbell bicep curl (isotonic)/ dumbell hold at 90 $^{\circ}$ of flexion(isometric)}                  \\ \hline
Weight(kg)             &  No weight \vline  & 1kg \vline &  2.5kg \vline & 5kg \vline  & 6kg \vline & 9kg \vline & 10kg \\ \hline
Muscles        & \multicolumn{7}{c}{BB and FCU}  \\ \hline
Subjects       & \multicolumn{7}{c}{9}                  \\ \hline
Rest duration          & \multicolumn{7}{c}{10s}                              \\ \hline
Activity duration      & \multicolumn{7}{c}{8s}                              \\ \hline
No of repetitions        & \multicolumn{7}{c}{09}                               \\  \hline
sEMG sensor            & \multicolumn{7}{c}{Noraxon}                          \\  \hline
Electrode              & \multicolumn{7}{c}{silver(Ag)/silver chloride(Agcl)} \\ \hline
Sampling frequency(Hz) & \multicolumn{7}{c}{2000}                             \\ \hline
No of channels         & \multicolumn{7}{c}{2}  \\
\hline \hline
\end{tabular} 
}
\label{tab:EMAHA-DB2}
\end{table}

\begin{table}[htb]
\centering
\caption{DETAILS OF PARTICIPANTS } 
\resizebox{\columnwidth}{!}{
\begin{tabular}{cccccc}
\hline \hline
\textbf{Subject} & \textbf{Circumference of BB muscle (inches)} & \textbf{Circumference of forearm (inches)} & \textbf{Experience}             & \textbf{Weight (kg)}              & \textbf{Height (cms)}                  \\ \hline
\textbf{1}       & {\color[HTML]{000000} 10.5}                  & {\color[HTML]{000000} 10}                  & No                             & {\color[HTML]{000000} 58} & {\color[HTML]{000000} 175}                                 \\ \hline
\textbf{2}       & {\color[HTML]{000000} 11.5}                  & {\color[HTML]{000000} 10.5}                & No                              & {\color[HTML]{000000} 75} & {\color[HTML]{000000} 183}                                  \\ \hline
\textbf{3}       & {\color[HTML]{000000} 13}                    & {\color[HTML]{000000} 10.5}                & No                              & {\color[HTML]{000000} 70} & {\color[HTML]{000000} 173.7}   \\ \hline
4.               & {\color[HTML]{000000} 12.8}                  & {\color[HTML]{000000} 10.5}                & {\color[HTML]{000000} 1 month}  & {\color[HTML]{000000} 81 } & {\color[HTML]{000000} 182}    \\ \hline
5                & {\color[HTML]{000000} 11}                    & {\color[HTML]{000000} 9.8}                 & {\color[HTML]{000000} 1 month}   & 57                         & 174                            \\ \hline
6                & {\color[HTML]{000000} 12}                    & {\color[HTML]{000000} 10}                  & {\color[HTML]{000000} 2 months} & 75                         & 182                           \\ \hline
7                & 13.8                                         & 11.9                                       & {\color[HTML]{000000} 1 year} & {\color[HTML]{000000} 65 } & {\color[HTML]{000000} 173 }    \\ \hline
8                & {\color[HTML]{000000} 14.3}                  & {\color[HTML]{000000} 12}                  & {\color[HTML]{000000} 2 years}  & {\color[HTML]{000000} 77 } & {\color[HTML]{000000} 176 }    \\ \hline
9                & {\color[HTML]{000000} 13.8}                  & {\color[HTML]{000000} 12.2}                & {\color[HTML]{000000} 1 year}   & {\color[HTML]{000000} 79 } & {\color[HTML]{000000} 182.8 } \\ \hline \hline
\end{tabular}
}
\label{details_participants}
\end{table}

\section{Results}
\subsection{Performance Comparisons}
\textcolor{black}{To validate the suitability of the LGM model, the fit of the proposed model to the sEMG data is evaluated in comparison with the following models.}
\begin{itemize}
    \item LGM model (proposed).
    \item Standalone Laplacian (SL) model \cite{hussain2009electromyography}.
    \item Standalone Gaussian  (SG) model \cite{roesler1974statistical}.
    \item Scale-mixture (SM) model \cite{furui2019scale}.
\end{itemize}

\subsubsection{Evaluation methods}  \label{Em}
The following qualitative and quantitative analyses are used to evaluate the suitability of the LGM model.

\begin{itemize}

    \item \textcolor{black}{
    Visual comparisons\cite{kusuru2021laplacian}:{ The LGM model is visually compared with the empirical distribution (mpdf) and the process is repeated for the other three models.}}
    \item 
    \textcolor{black}{{Area Difference (AD)\cite{clancy1999probability}}:
    It is a statistical metric to determine how much the empirical pdf differs with the proposed theoretical pdf.} It is defined as follows
\begin{equation}
    AD=\sum_{l=1}^{K} \left | E(l)-P(l) \right |
\end{equation}
where $E$ and $P$ correspond to the empirical and proposed models respectively.
    
    \item 
    \textcolor{black}{
    Kullback–Leibler divergence (KLD)\cite{kusuru2021laplacian}: KLD evaluates the difference between
two probability distributions. It can be written as 
\begin{equation}
    KLD(p_1,p_2)=E\left [ ln \bigg(\frac{p_{1}(x)}{p_{2}(x)}\bigg) \right ]=\sum_{x} p_1(x) ln\left [ \frac{p_{1}(x)}{p_{2}(x)} \right ]
\end{equation}
\textcolor{black}{here $p_1$ corresponds to LGM model and $p_2$ corresponds to existing models}. The lower the values of the KLD and AD, the better is the model fit.  }

    \item Likelihood Ratio test (LRT)\cite{methodology1989likelihood}:

\textcolor{black}{
It is based on the logarithm of ratio of the likelihoods of two competing models with the unknown parameters estimated form the data using the maximum likelihood estimation. The test statistic is compared against a specific 
threshold and the corresponding hypothesis is accepted or rejected accordingly.  }
\end{itemize}

\subsubsection{Visual comparisons}
\begin{figure*}[!t]
\centering
\begin{subfigure}[t]{0.24\textwidth}
    \includegraphics*[width=1.05\columnwidth]{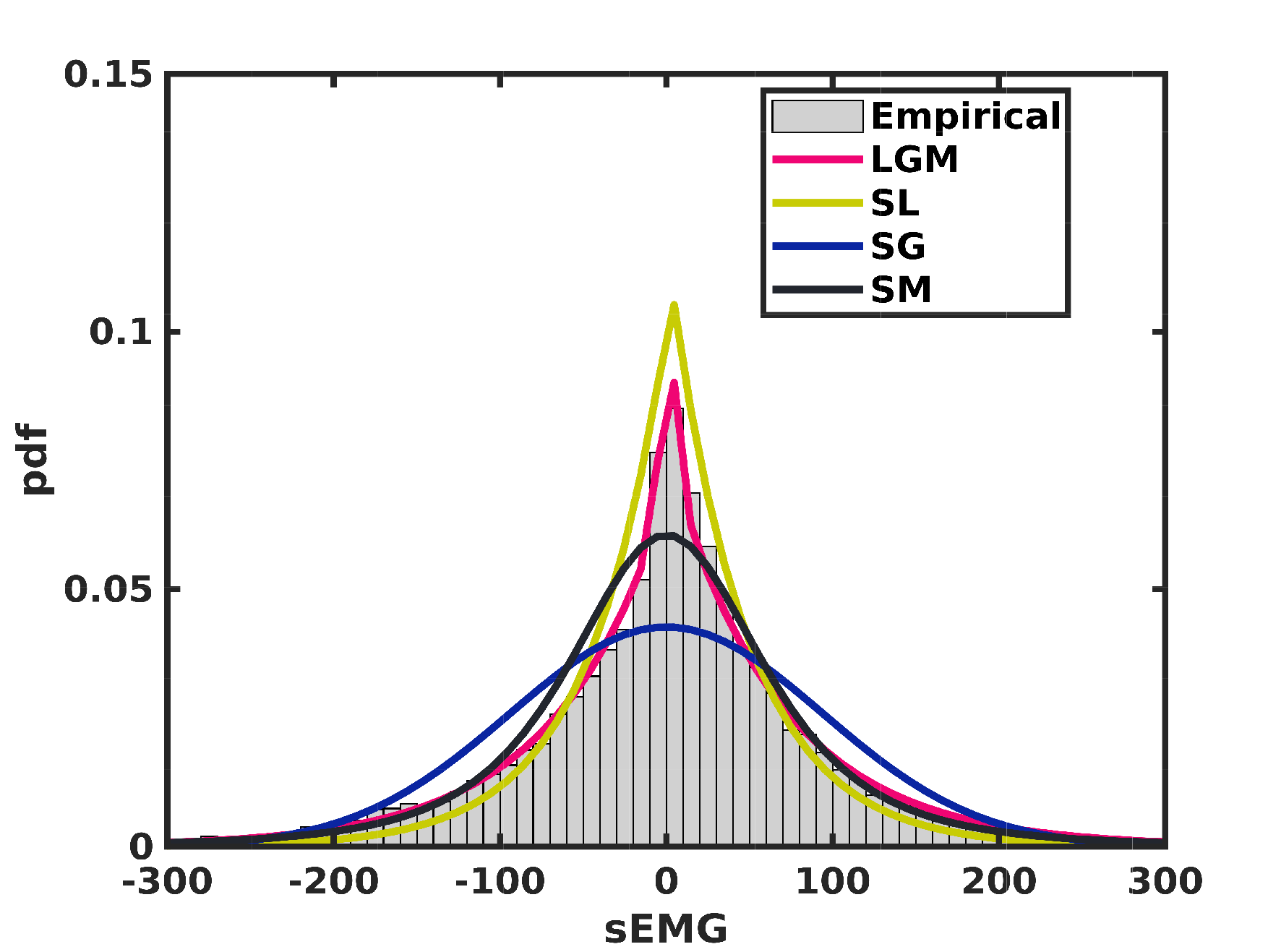}
    \caption{BB}
       \label{isotonic_BB}
\end{subfigure}%
    ~
\begin{subfigure}[t]{0.24\textwidth}
    \includegraphics*[width=1.05\columnwidth]{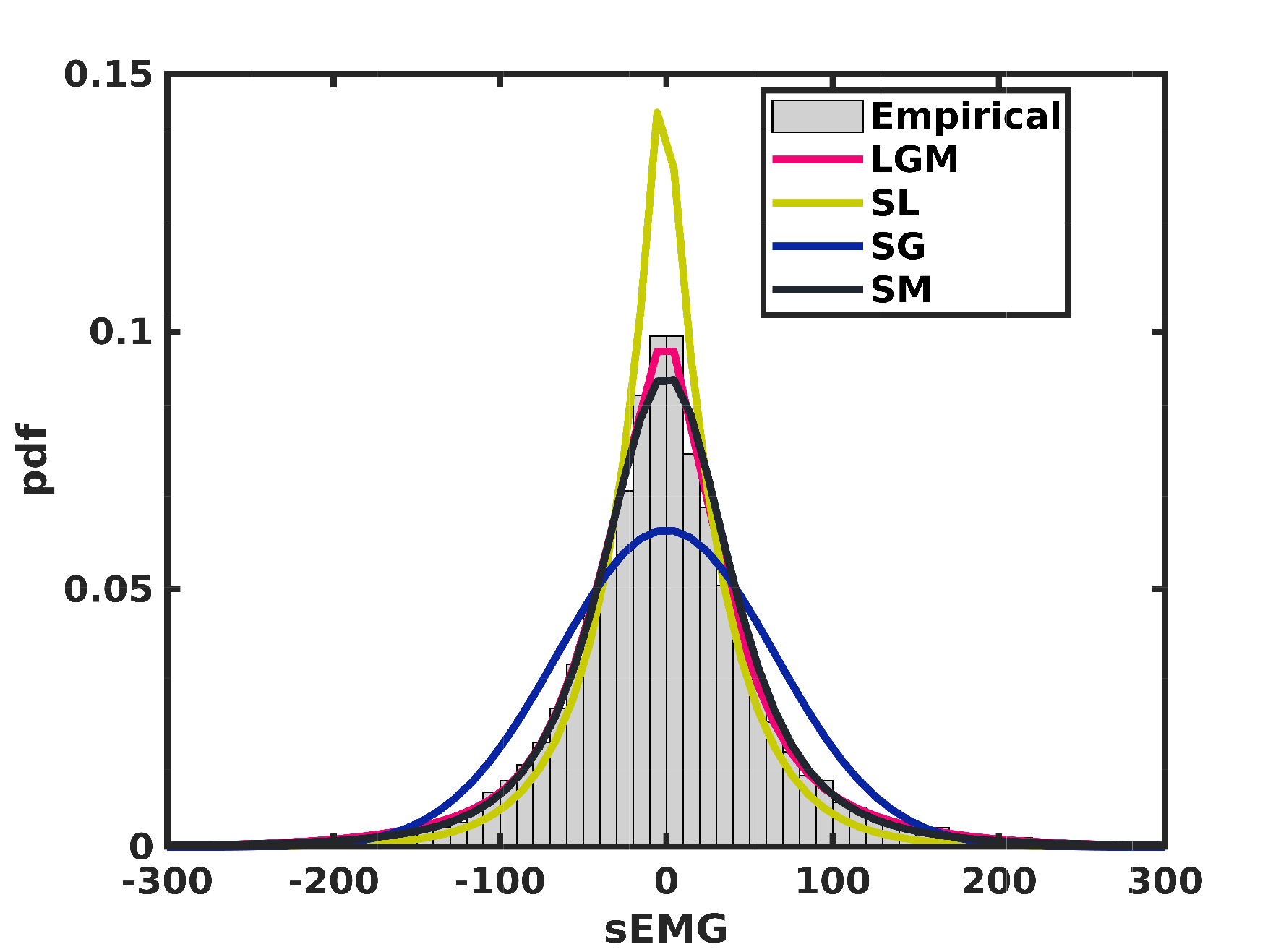}
    \caption{FCU}
    \label{isotonic_FCU}
\end{subfigure}%
    ~
\begin{subfigure}[t]{0.24\textwidth}
    \includegraphics*[width=1.05\columnwidth]{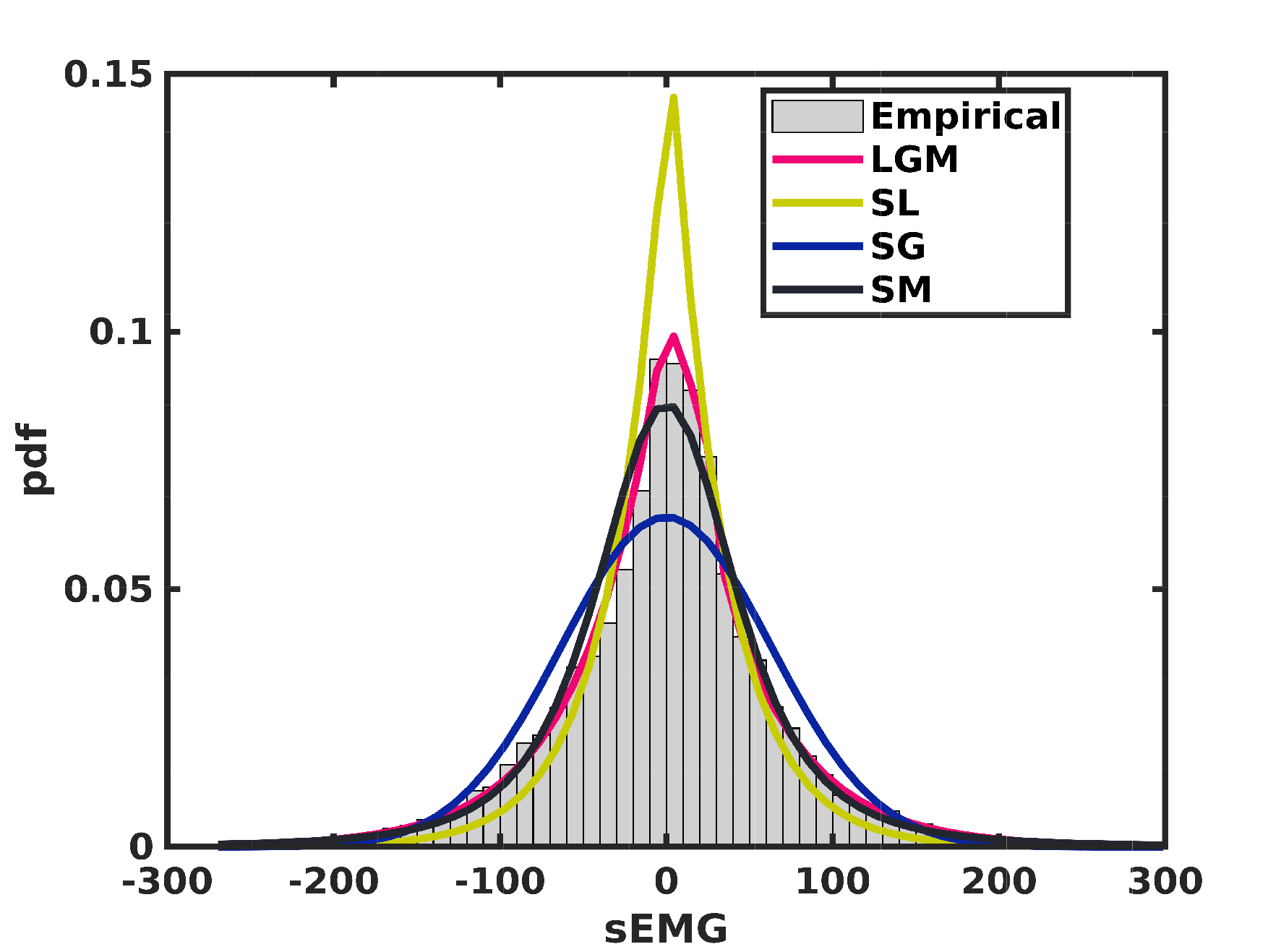}
    \caption{BB}
   \label{isometric_BB}
\end{subfigure}
\begin{subfigure}[t]{0.24\textwidth}
    \includegraphics*[width=1.05\columnwidth]{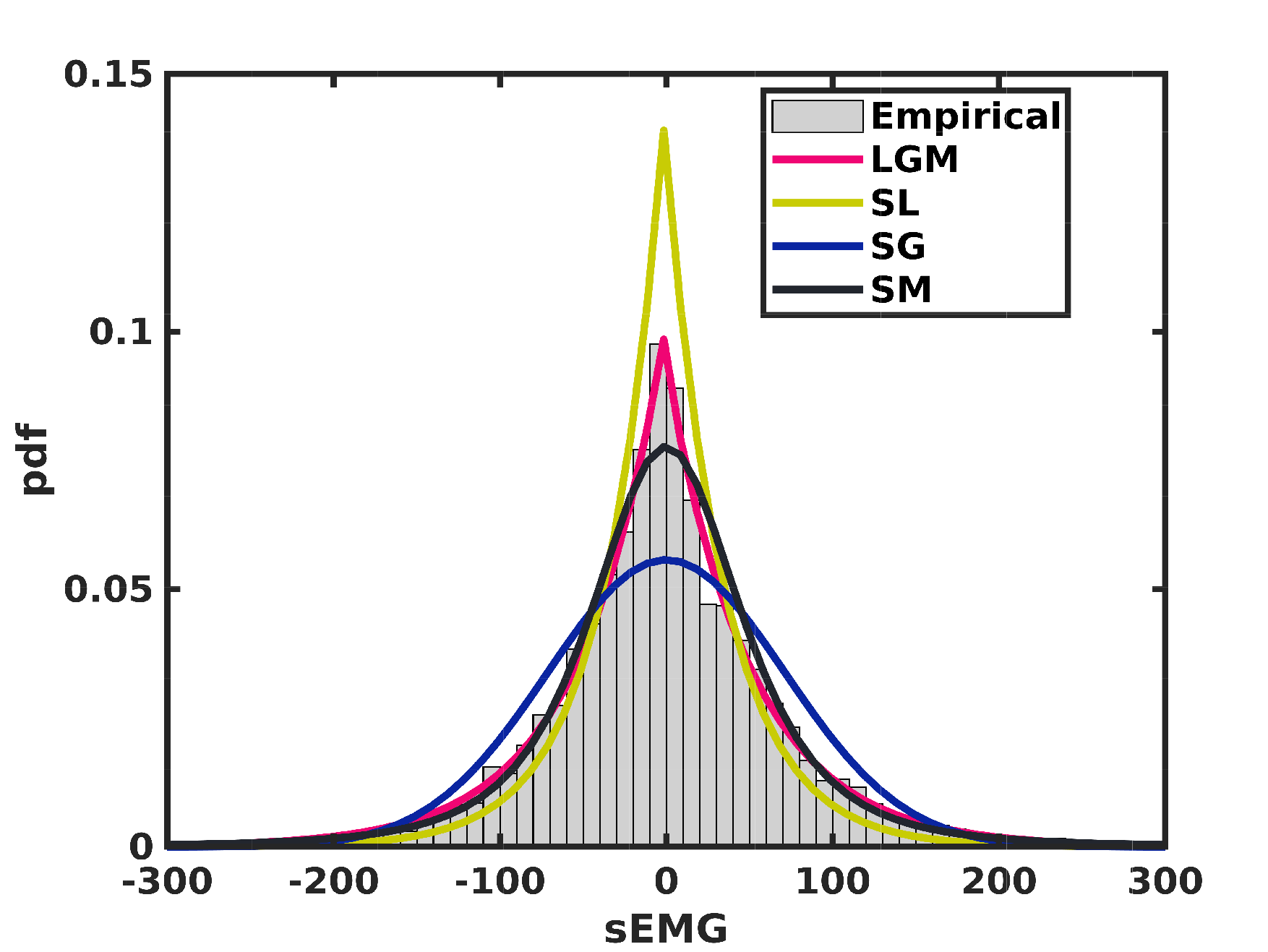}
    \caption{FCU}
    \label{isometric_FCU}
\end{subfigure}%
\caption{Visual comparison between histogram-based model (gray) with LGM (Pink), Laplacian (yellow), Gaussian (blue) and SMM (black) models for a subject-$7$ with a weight-$2$kg. (a) and (b) correspond to weight training during isotonic activities. (c) and (d) correspond to weight training during isometric activities.} 
\label{VC}
\end{figure*}

\textcolor{black}{
Fig.\ref{VC} illustrates the mpdf and estimated pdfs of the LGM model and other existing models for subject-$7$, with a $2.5$kg dumbell. Specifically, Fig.\ref{VC} (a) shows pdfs for isotonic activity at BB, Fig. \ref{VC}(b) shows pdfs for isotonic activity at FCU, Fig. \ref{VC}(c) shows pdfs for isometric activity at BB and Fig. \ref{VC}(d) shows pdfs for isometric activity at FCU. These pdfs clearly demonstrate that among the four models, the LGM model has higher coverage of the mpdf. The same analysis is carried for the rest of trials, activites, and subjects. It is observed that the LGM model has the best agreement with the mpdf.}

\begin{figure*}[t!]
\centering
\begin{subfigure}[t]{0.45\textwidth}
    \includegraphics*[width=1\columnwidth]{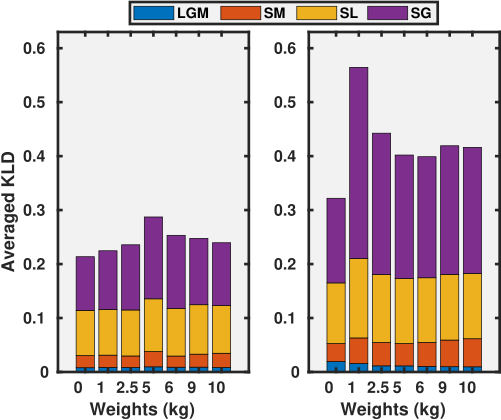}
    \caption{}
     \label{KLD_isotonic_BB}
\end{subfigure}%
    ~
\begin{subfigure}[t]{0.45\textwidth}
    \includegraphics*[width=1\columnwidth]{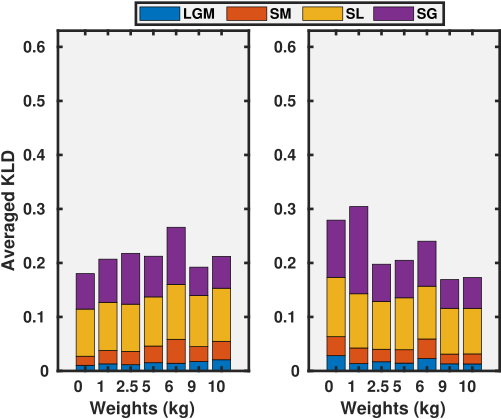}
    \caption{}
    \label{KLD_isotonic_FCU}
\end{subfigure}%
\caption{Comparison of average KL-divergence across trials and subjects with LGM, SMM, Laplacian and Gaussian models: (a) BB (left) and FCU (right) corresponding to isotonic contractions (b) BB (left) and FCU (right) corresponding to isometric contractions.}
\label{KLD}
\end{figure*}

\subsubsection{KLD Maps and Plots}
\textcolor{black}{
For a given subject, for each activity and trial, the KLD is evaluated between the mpdf and a model pdf for eg. the LGM pdf.  Fig \ref{KLD} shows the average KLD of LGM model and average KLDs from other existing models. For a given dumbbell weight, averaging is carried out across the trials and subjects. The horizontal axis shows the weights  and the vertical axis shows the average KLD. Fig \ref{KLD}(a) corresponds to isotonic contractions during weight training activities from BB (left) and FCU (right). Fig \ref{KLD}(b) corresponds to isometric contractions from the same muscles. From these figures it is evident that the average KLD is lowest in the case of LGM model compared to the SM and other standalone models.  From Fig. \ref{KLD} it is observed that the KLD is lower in the case of isotonic contractions when compared to isometric contractions. Moreover, the average KLD values consistently increase in the following order: LGM, SM, SL, and SG.  This means among the four models, the standalone Gaussian is the least fit and the LGM is the best fit.
}

\begin{figure*}[!t]
\centering
\begin{subfigure}[t]{0.45\textwidth}
    \includegraphics*[width=1\columnwidth]{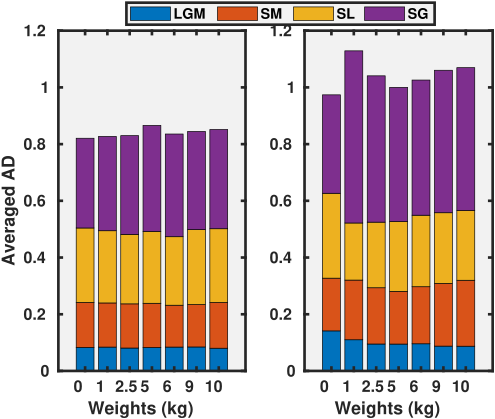}
    \caption{}
     \label{AD_isotonic_BB}
\end{subfigure}%
    ~
\begin{subfigure}[t]{0.45\textwidth}
    \includegraphics*[width=1\columnwidth]{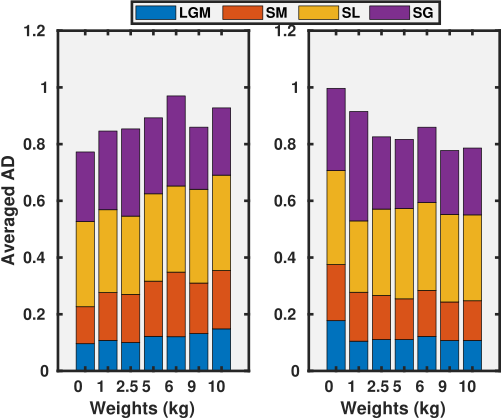}
    \caption{}
    \label{AD_isotonic_FCU}
\end{subfigure}%
\caption{Comparison of average Area Difference across trials and subjects with LGM, SMM, Laplacian and Gaussian models: (a) BB (left) and FCU (right) correspond to isotonic contractions; (b) BB (left) and FCU (right) correspond to isometric contractions}
\label{AD}
\end{figure*}

\subsubsection{Area difference (AD)}
\textcolor{black}{
Again, for each subject, each activity and each trial, the AD is computed between the estimated model pdf and the mpdf. Fig. \ref{AD} depicts AD averaged across subjects and trials between the LGM and the mpdf.  It also includes the average AD between the mpdf and the corresponding pdfs from the SM and the other standalone models. The horizontal axis represents the weights and the vertical axis represents the averaged AD. Fig. \ref{AD} (a) shows the averaged AD at the BB (left) and the FCU (right) corresponding to the isotonic contractions.  Fig. \ref{AD} (b) shows the averaged AD at BB and FCU corresponding to the isometric contractions.  From Fig \ref{AD}, for the isotonic and isometric contractions, the averaged AD is lowest for the LGM model compared to the other three models. Again, under the isotonic contractions the average AD for the LGM is $10$ percent better than the averaged AD for the SM. Further, in the case of isometric contractions, it is $8$ percent better than that of the SM. Thus, based on averaged AD, the SM model is second best fit after the LGM followed by the SL and the SG has the least fit.}

\subsubsection{LRT results}
LRT is carried out between LGM model and other existing models. Let $H_o$ and $H_1$ denote null hypothesis and alternative hypothesis as existing models and the LGM model fit to the sEMG signals respectively. 
\begin{eqnarray}
H_0: & \text{Existing models fit EMG signals} \nonumber \\
H_1: & \text{LGM model fits EMG signals}   \nonumber 
\end{eqnarray}
Using $95\%$ confidence intervals, the p-value is $0.028$. Since it is less than $0.05$ the $H_0$ should be rejected. Until this subsection, the focus is on the sEMG signals and comparisons between the LGM  and the other existing models.

\begin{figure*}[!t]
     \centering
\begin{subfigure}[t]{0.45\textwidth}
      \includegraphics*[width=.9\columnwidth]{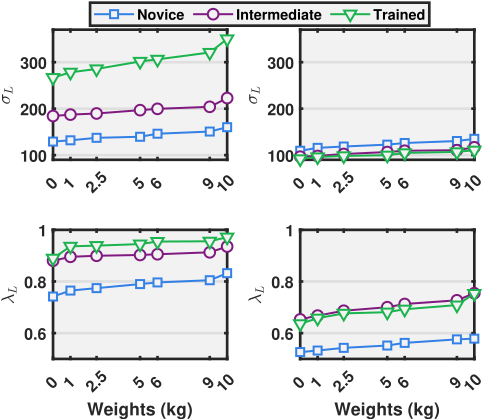}
       \label{Experience_ISOTONIC}
     \caption{Isotonic activity}
\end{subfigure} ~
\begin{subfigure}[t]{0.45\textwidth}
     \includegraphics*[width=0.9\columnwidth]{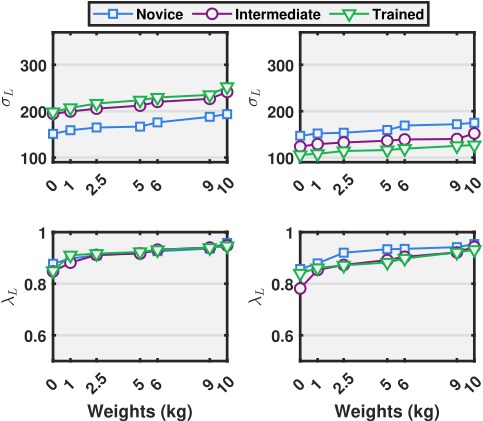}
       \label{Experience_ISOTONIC}
     \caption{Isometric activity}
\end{subfigure}
   \caption{Standard deviations (top)  of the Laplacian component at the  BB  (left) and  FCU  (right) and 
       Laplacian mixture weights (bottom) corresponding to  BB  (left) and  FCU  (right) as function of lifted weights for different training experiences}
     \label{Experience_sig}
\end{figure*}

\begin{figure*}[!t]
     \centering
\begin{subfigure}[t]{0.45\textwidth}
      \includegraphics*[width=.9\columnwidth]{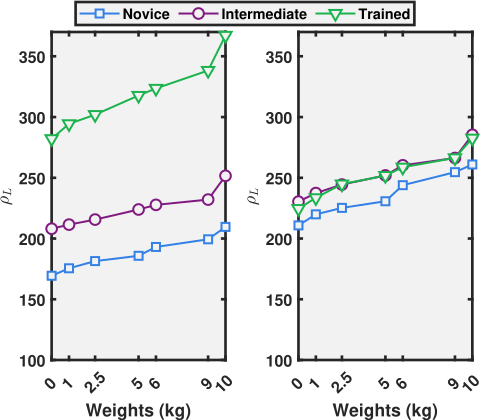}
       \label{subfig:rho}
     \caption{Square root of L-power $\rho^2_L$}
\end{subfigure} ~
\begin{subfigure}[t]{0.45\textwidth}
     \includegraphics*[width=0.9\columnwidth]{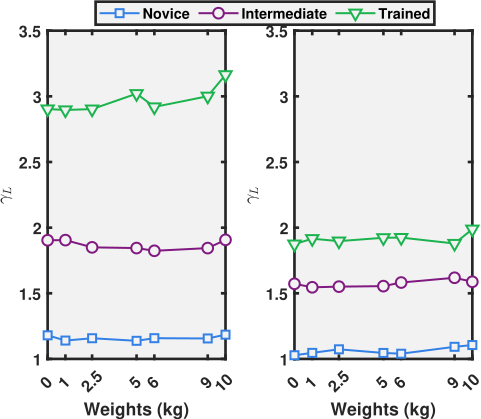}
       \label{subfig:gam}
     \caption{Ratio $\gamma_L$}
\end{subfigure}
   \caption{Characteristics of standard deviations of laplacian components from BB and FCU for Isotonic (left) and Isometric (right) activities}
     \label{fig:Rho_Gam}
\end{figure*}
 
\section{Analysis and Discussion}
\subsection{Impact of measurement factors on Laplacian parameters}
\textcolor{black}{In this section, the focus is on analysis of parameters of the LGM model and the impact of various measurement conditions on them. Specifically, the conditions considered are 1) training experience, 2) weights, 3) muscle type, and 4) contraction type. 
}
\subsubsection{Impact on Laplacian standard deviation $\sigma_L$ and mixing weights $\lambda_L$}
\textcolor{black}{
Figs \ref{Experience_sig}(a) and \ref{Experience_sig}(b) represent the variations of standard deviations ($\sigma_L$) and the Laplacian weights ($\lambda_L$) of the Laplacian component in the LGM model as a function of lifted weights for different training experiences (Novice (blue), Intermediate (purple), and trained (green) during isotonic and isometric contractions. The parameters $\sigma_L$  and $\lambda_L$ are averaged out across all trials and subjects. In Figs \ref{Experience_sig}(a) and \ref{Experience_sig}(b), the top and bottom rows correspond to $\sigma_L$ and $\lambda_L$ respectively. It is interesting to note that both $\sigma_L$ and $\lambda_L$ show a rising trend with respect to dumbbell weights. An increasing $\sigma_L$ confirms that force required to train with dumbbell increases with its weight and an increasing $\lambda_L$ suggests that the contribution of the Laplacian components increases with the load. 
Further, $\sigma_L$ from BB during both contractions are higher than those from the FCU. This observation suggests that the BB generates greater force compared to the FCU for both the training activities. Further, in the isotonic case, at BB the $\sigma_L$ increases consistently with the subject's experience. Thus for bicep curls, the muscle force generated by BB increases with training experience. Whereas in isometric case, the $\sigma_L$'s do not differ significantly with experience.}
\subsubsection{Impact on root of sum of variances}
\textcolor{black}{
The metric $\rho_L$ denotes the square root of sum of laplacian variances from the two muscle sites.
\begin{equation}
    \rho_L = \sqrt{\sigma^2_{L,BB} + \sigma^2_{L,FCU}}   
\end{equation}
The $\rho^2_L$ termed as "L-power" is the Laplacian component of the sEMG signal power and can be directly related to the muscle force. 
Fig. \ref{fig:Rho_Gam}(a) depicts $\rho_L$ from BB and FCU for Isotonic (left) and Isometric (right) activities. For isotonic activity, specifically dynamic bicep curls, for a any lifting load, the signal power (L-power) $\rho^2_L$  seems to be directly related to the subject's experience. For any weight, the trained subjects produced the highest L-power while the novices generated the least L-power. For isometric activity, specifically static dumbbell holding for any load, the  L-power $\rho^2_L$  does not increase significantly with  subject's experience i.e., these trends are statistically similar. It is validated using the F-test \cite{johnston1963econometric} with $5\%$ significance level. This statistical behavior can be explained by the nature of muscle activity. For instance, during the isotonic activity specifically during bicep curls, the muscle forces generated need to exceed the lifting weight for a proper lift and the "trained" subjects can produce greater muscle force. However in the case of isometric activity (dumbbell hold) the muscle forces need to match the loading weight, in other words an equilibrium is sufficient. Thus the L-power need not increase significantly with training experience.
}
\textcolor{black}{
In addition, it is interesting that $\rho_L$ can be correlated with a subject's strength. A higher slope for $\rho_L$ vs weights 
indicates higher strength to lift heavier weights. From fig. \ref{fig:Rho_Gam}(a), for isotonic activity, the trained group has a steeper slope of $\rho_L$, followed by the intermediate and the novice groups.}

\subsubsection{Impact on ratio of Laplacian standard-deviations}
\textcolor{black}{
The ratio $\gamma_L$ is defined as 
\begin{equation}
    \gamma_L =  \frac{\sigma_{L,BB}}{\sigma_{L,FCU}}  
\end{equation}
Fig. \ref{fig:Rho_Gam}(b) shows $\gamma_L$ for Isotonic (left) and Isometric (right) activities. It is interesting to note that for any activity (isometric or isotonic), the ratio $\gamma_L$ has nearly a fixed value as a function of lifting weights and is an increasing function of the training experience. 
The ratio $\gamma_L$ indicates how the signal power is distributed between the two muscle sites. Clearly the signal power generated depends on the muscle site,
the type of activity and the training experience of the subject. 
} 

\subsection{Discussion}
 \subsubsection{Characteristics of the LGM}
  \textcolor{black}{
Based on the qualitative and quantitative analysis, the LGM model seems to fit best to the sEMG signal strength compared to the SM and the standalone models. Further, for various measurement conditions such as muscle sites, type of activity, loading weight and experience of subjects the LGM model is the best fit. Additionally, the model parameters show dependencies on these conditions. Observations from these analyses can be summarized as 1) both the Laplacian and Gaussian components contribute to the model and the mixing weights depend on the type of muscle, type of muscle contraction (activity), and subject's training experience. Generally, the Laplacian contribution is stronger than the Gaussian component. This is consistent with the findings reported by Hussian et al \cite{hussain2009electromyography,kaplanis2000bispectral}. The standalone Laplacian model alone is unable to fit the sEMG signal pdf because the Gaussian component also contributes 2) the sEMG signal's L-power $\rho^2_L$ increases with weights and in the isotonic contractions case, it increases with experience as well. 
}

\subsubsection{Relation to muscle force distribution}
\textcolor{black}{
Biomechanics of the arm indicates that the muscle force generated is proportional to the loading weight. From the sEMG literature it is evident that the sEMG signal power is an indicator of the force generated. Thus $\gamma_L$ can be an indicator of how the work load (generation of muscle force) is distributed between the muscle groups. For example, in the case of novice subjects, both the biceps and forearm muscles generate nearly equal portions. As the experience increases, the biceps take up major portion of the work load and the forearm muscles only need to provide mechanical support. Importantly, the work load distribution seems to be nearly independent of weight lifted and adapts with the training experience and the type of activity. Additionally, the muscle force is directly related to muscle fiber recruitment. It can be inferred that fiber recruitment increases with lifted weights. Further, for an activity as biceps curls, for subjects with more training experience, the fiber recruitment may get specialized to one of the muscle sites (BB) and reduces the load on fibers of the FCU.}

\section{Conclusion}
\textcolor{black}{
In this study, we collected a new sEMG dataset (EMAHA- DB2) for the Indian population. 
The sEMG signals were acquired from the BB and the FCU during isometric and the isotonic activities with dumb-bells. The suitability of the LGM model is analysed for these sEMG signals strength as a function of loading weights. Based on visual comparison, KL-divergence, Area difference, and LRT test, the LGM model performs better compared to the existing SMM and standalone models. The LGM model was found to be the best fit to the mpdf under various measurement conditions such as type of muscle, type of activity, lifted weight and the training experience of a subject. Based on  variance of the Laplacian component from the sEMG signal, the force generated by the 
arm directly depends on both the weight lifted and the subjects training experience. Further the ratio
of the forces generated by the muscle groups (BB and FCU) is found to be relatively fixed with respect to the weight lifted and increases consistently with the training experience. {Future plans include extension to multiple channel EMG data, an analysis of the LGM model and its parameters with respect to conditions such as muscle fatigue and different activities of daily living.} }

\section*{Acknowledgment}
This research is funded by SERB, Govt. of India under Project Grant No. CRG/2019/003801.

\bibliographystyle{IEEEtran}
\bibliography{ref.bib}

\end{document}